\documentclass{mem}
\usepackage{natbib}\usepackage{txfonts}\usepackage{balance}
\usepackage{graphicx}
\usepackage[a4paper,breaklinks,dvipdfm]{hyperref}
\idline{75}{282}
\begin{document}

\newcommand{\Halpha}{H$\alpha$}
\newcommand{\kms}{km\,s$^{-1}$}
\newcommand{\ms}{m\,s$^{-1}$}

\newcommand{\hhh}{\mbox{H$_2$ }}
\newcommand{\hhhns}{\mbox{H$_2$}}
\newcommand{\ma}{\mbox{m\AA}}
\newcommand{\dmu}{\mbox{$\Delta\mu/\mu$ }}

\title{
Constraints of \dmu based on \hhh observations in QSO spectra
at high redshifts}

   \subtitle{}

\author{
M.~Wendt\inst{1} 
\and P.~Molaro\inst{2}
          }

  \offprints{M. Wendt}

\institute{
Institute of Physics and Astronomy,
University Potsdam, 14476 Potsdam,
Germany\\
\email{mwendt@astro.physik.uni-potsdam.de}
\and
Intituto Nazionale di Astrofisica --
Osservatorio Astronomico di Trieste, Via Tiepolo 11,
I-34131 Trieste, Italy
}

\authorrunning{Wendt}

\titlerunning{Constraints of \dmu}

\abstract{
This report summarizes the latest results on the 
proton-to-electron mass ratio $\mu$ we obtained  from \hhh observations at high redshift 
in the light of possible variations of fundamental physical constants.
The focus lies on a better understanding of the general error budget that led to disputed
 measurements of \dmu in the past. Dedicated observation runs, and alternative approaches to 
improve 
 accuracy provided results which are in reasonable good agreement with no 
variation and provide an upper limit of $\left| \dmu \right|$ $<  1\times 10^{-5}$ for the redshift
range of $2 <$ z $< 3$.
}
\maketitle{}

\section{Introduction}
Our Standard Model contains numerous
fundamental physical constants whose values cannot be predicted by theory and   
therefor need to  be measured through experiments \citep{Fritzsch09}.
 These are mainly the masses of the elementary particles and the  dimensionless 
coupling constants.
The latter are assumed to be time-invariant  although   
 theoretical models    which seek  to unify
the four forces of nature usually allow them to vary naturally on cosmological
scales. 
 The proton-to-electron mass ratio, $\mu  = m_{\mathrm{p}} / m_{\mathrm{e}} = 1836.152 672 45(75)$
\footnote{2010 CODATA recommended value.} and the 
fine-structure constant $\alpha \equiv e^2/(4\pi \epsilon_0 \hbar c) \approx 1/137$  
are two specific constants that can be probed in the laboratory as well as  in the  
distant and early Universe.
Observations of  absorption lines in the spectra of intervening systems
 towards  distant quasars (QSO) have  been 
 subject of numerous studies. 

The fine-structure constant is related to the electromagnetic force while $\mu$
is  sensitive primarily   to the quantum chromodynamic scale \citep[see, i.e.,][]{Flambaum04}. 
The $\Lambda_{{\mathrm{QCD}}}$  scale
is supposed to vary considerably faster than that of quantum electrodynamics
$\Lambda_{{\mathrm{QED}}}$. Consequently, the change in the proton-to-electron mass ratio,
 if any, is expected to be larger than that of the fine structure constant.   
Hence,\,\,$\mu$  is an ideal candidate to search for possible cosmological
variations of the fundamental constants.

A measure of $\mu$ can for example be obtained by comparing
relative frequencies of the electro-vibro-rotational lines of \hhh
as first applied by \citet{Varshalovich93}
 after \citet{Thompson75}
 proposed the general approach to utilize molecule transitions for $\mu$-determination.
\begin{figure}
\includegraphics[bb=50 50 410 320,width=\linewidth]{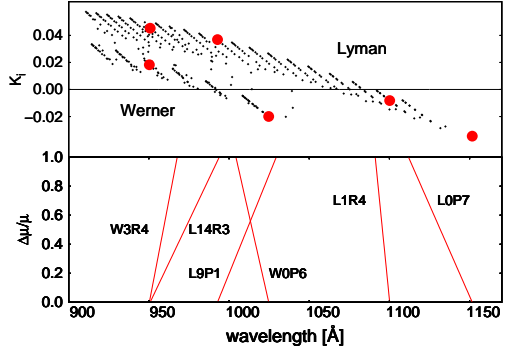}
\caption{The upper panel shows the sensitivity coefficients $K_i$ of the Lyman and Werner
transitions of \hhh against the restframe wavelength. Note, that the coefficients show 
different signs. The lower panel demonstrates the shifts of six selected transitions 
(marked with large red circles in the upper panel) with increasing \dmu. 
For this illustration the range of \dmu is five orders of magnitude larger
than the current constraints on \dmu.}
\label{fig:ki}
\end{figure}
The applied method uses the fact that the wavelengths of vibro-rotational
lines of molecules depend on the reduced mass, M, of the molecule.
Comparing electro-vibro-rotational lines with different
sensitivity coefficients gives a measurement of $\mu$.

The observed wavelength $\lambda_{\mathrm{obs},i}$  of any given
line in an absorption system at the redshift $z$ differs from the local
rest-frame
wavelength $\lambda_{0,i}$  of the same line in the laboratory according to the
relation 
\begin{equation}\label{eq:1}
\lambda_{\mathrm{obs},i}  =  \lambda_{0,i} (1 + z)\left(1+ K_i \frac{\Delta \mu}{\mu}\right),
\end{equation}   
where $K_i$ is the sensitivity coefficient of the $i$th component computed
theoretically for the Lyman and Werner bands of the \hhh
molecule \citep{Meshkov07,Ubachs07}.
Figure \ref{fig:ki} shows the sensitivity coefficients $K_i$ for the Lyman and Werner
transitions of \hhh in the upper panel. The coefficients are typically on the order
of $10^{-2}$. Since several coefficients differ in sign, some \hhh lines are 
shifted into opposite directions in case of a varying $\mu$.
The lower panel demonstrates this effect.
The corresponding sensitivity coefficients are marked as {\it filled red circles} in the upper panel.
The expected shifts at the current level of the constraint on \dmu are on the order of a 
few \mbox{100 \ms} or about 1/10$^{\mathrm{th}}$ of a pixel size.

Line positions are usually given as relative velocities with comparison to the redshift
 of a given absorption system defined by the redshift position of the lines with $K_i \approx0$,
 then introducing the reduced redshift  $\zeta_i$:
\begin{equation}
\zeta_i \equiv \frac{z_i - z}{1+z}
 = K_i \frac{\Delta\mu}{\mu}. \label{eq_LBLFM}
\end{equation}
 The velocity shifts of  the lines are   linearly proportional
 to $\Delta\mu/\mu$ which  can be measured  through a regression analysis in the
 $\zeta_i - K_i $ plane. 
This approach is referred to as line-by-line analysis in contrast to the 
comprehensive fitting method (CFM) which will be discussed in 
section \ref{sec:risks}.

\section{\dmu at the highest redshift}
One of the latest results for \dmu is described in
\citet{Wendt12} and based on observations of
 QSO 0347-383. The damped Lyman-$\alpha$ system (DLA) at $z_{\mathrm{abs}}=3.025$ 
in its spectrum bears many absorption features of molecular hydrogen and represents the
\hhh system with the highest redshift utilized for \dmu measurements.
Numerous different results on \dmu in the redshift range of $2 <$ $z_\mathrm{abs} < 3$
 led to the conclusion
that the wavelength calibration had become the limiting factor in constraining \dmu.
Data of unprecedented quality in terms of resolution and calibration exposures was required.
This and other aspects motivated the
 ESO Large Programme\footnote{ESO telescope programme L185.A-0745} and several
data taken in advance to verify the spectrograph setup expected to provide the best results.
The data of QSO 0347-383 were taken with the Ultraviolet and Visual Echelle Spectrograph (UVES)
at the Very Large Telescope (VLT) on the nights of September 20-24 in 2009.
The CCD pixels were not binned for these exposures for maximum resolution.
A pixel size  of $0.013-0.015$ \AA, or $1.12$ \kms\  at 4000 \AA\, along the dispersion direction
was achieved.
 More details are given in \citet{Wendt12}.
\section{Results for QSO 0347-383}
\begin{figure}
\includegraphics[clip=true, width=\linewidth]{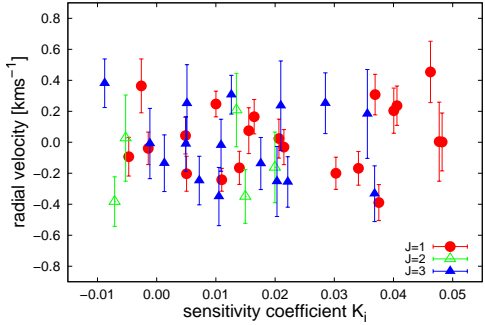}
\caption{Measured radial velocity vs. sensitivity for 42 \hhh 
lines observed in QSO 0347-383 in \citet{Wendt12}. 
Any correlation would indicate a change in $\mu$. The different symbols  correspond
to the three observed rotational levels. The errorbars reflect the 1 $\sigma$ errors.}
\label{fig:v_ki}
\end{figure}
In Figure \ref{fig:v_ki} the  measured radial velocities of the 42 \hhh lines measured in QSO 0347
are plotted against the sensitivity  coefficients $K_i$ of the corresponding transition. 
Any correlation therein would indicate a variation of $\mu$ 
at $z_{\mathrm{abs}}=3.025$ with respect to laboratory values. 
The data give no hint towards variation of the proton-to-electron
 mass ratio in the course of cosmic time.
The uncertainties  in the line positions of the \hhh features due to
 the photon noise are  estimated by the fitting algorithm.
 These are shown in the errorbars
in Figure \ref{fig:v_ki}.
 The mean error  in the line positioning is of \mbox{150 \ms}.
 Even at first glance the given errorbars in Figure \ref{fig:v_ki} appear to be too
small to explain the observed scatter. 

The scatter of lines with similar sensitivity  coefficients $K_i$ directly reflects
the uncertainties in the line positions as it cannot be attributed to possible variations of $\mu$
since it is present for basically the same sensitivity parameter. The intrinsic
scatter is of the order of 210 \ms and thus larger than the positioning error of the individual
lines.
That is also reflected by a reduced $\chi^2$ of 2.7 for a weighted linear fit to the
data  (corresponding to  $\Delta\mu/\mu = (1.8 \pm 8.2) \times
10^{-6}$ at $z_{\mathrm{abs}}=3.025$). The factual scatter of
the data of the order of 210 \ms constitutes an absolute limit of precision. 
The above mentioned errors of the
fitting procedure require an  additional systematic component to
explain the observed scatter:
$\sigma_{\mathrm{obs}} \sim \sqrt{\sigma_{\mathrm{pos}}^2 +
\sigma_{\mathrm{sys}}^2},$
with $\sigma_{\mathrm{obs}} \approx$ 210 \ms, $\sigma_{\mathrm{pos}} \approx$ 150 \ms, and
$\sigma_{\mathrm{sys}} \approx$ 150 \ms.
A direct linear fit to the unweighted data yields:
$\Delta\mu/\mu = (4.2 \pm 7.7) \times 10^{-6}$.
Bootstrap analysis is a robust approach to obtain an estimate of the
 underlying linear relation of the data in Figure \ref{fig:v_ki}
 and estimate an error based on the true intrinsic scatter of the data.
A gaussian fit to the bootstrap gives $\Delta\mu/\mu = (4.3 \pm 7.2) \times 10^{-6}$
and is in good agreement with the direct methods applied.

\section{Challenges and risks}
\label{sec:risks}
There are in principle two approaches to determine \dmu based on line centroid measurements
of \hhh.
For the analysis of QSO 0347-383 we applied a straight forward linear regression of the measured
 redshifts of individual
\hhh absorption features and their corresponding sensitivity coefficients as plotted in Figure \ref{fig:ki}.
 This approach is referred to as line-by-line (LBL) analysis in contrast to the 
comprehensive fitting method (CFM).

The CFM  fits all \hhh components along with additional H\,{\sc i}
 lines and introduces an artificially applied $\Delta\mu/\mu$ as free parameter in the fit. 
 The best matching  $\Delta\mu/\mu$ is then derived via the resulting $\chi^2$ curve.
 The CFM aims to achieve the lowest possible reduced $\chi^2$ via additional velocity components.
 In this approach, the information of individual transitions is lost because merely the
overall quality of the comprehensive model is judged. 
%

The validity of the LBL or CFM approach depends
mostly on the analyzed \hhh system. For example, the absorption in 
QSO 0347-383 \citep[see][]{Wendt12} has the particular advantage of comprising
merely a single velocity component, which renders observed transitions independent of each other
and allow for this regression method. This was also tested in
 \citet{Rahmani13} and \citet{King08}.
For absorption systems with two or more closely and not properly resolved velocity components 
 many systematic errors may influence distinct wavelength areas. 

\citet{Weerdenburg11}, for example, increased the number of velocity components 
as long as the composite residuals of several selected absorption lines differed from flat noise. 
The uncertainties of the oscillator strengths $f_i$ that are stated to be up to 50\%
in the same publication might, however, further affect the choice
 for additional velocity components. 
A similar effect can be traced back to the nature of the bright background quasar
which in general is not a point-like source. In combination with the potentially
small size of the absorbing clumps of \hhhns, we may observe saturated absorption profiles with 
non negligible residual flux of quasar light not bypassing the \hhh cloud 
\citep[see][]{Ivanchik10}.

As pointed out by \citet{King11}, for multi-component structures with overlapping velocity
components the errors  in the line centroids are heavily correlated and a simple $\chi^2$ regression
is no longer valid. The same principle applies for co-added spectra with relative velocity shifts.
The required re-binning of the contributing data sets introduces
 further auto-correlation of the individual 'pixels'. 

\begin{figure}
\includegraphics[ width=\linewidth]{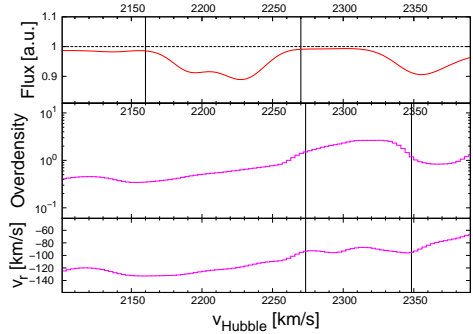}
\caption{Selected region from a simulated absorber as an example of a realistic density
distribution (\textit{center}) as well as macroscopic velocity fields (\textit{bottom}) leading to a multi-component
absorption feature (\textit{top}). The corresponding interval is marked with vertical lines.}
\label{fig:components}
\end{figure}

\citet{Rahmani13} discuss the assets and drawbacks of these two approaches 
in greater detail.
The selection criteria for the number of fitted components are non-trivial and under debate.
\citet{Prause13} discuss the possibility of centroid position
 shifts due to incorrect line decompositions
with regard to the variation of the finestructure constant $\alpha$,
 which in principle is applicable to
any high resolution absorption spectroscopy.
Figure \ref{fig:components} shows a small extract of data from extensive simulations as
an example of complex velocity and density structures that produce a multi-component
absorption profile.

Additionally, thermal-pressure changes  move in the cross dispersers in different ways,
 thus introducing relative shifts between
the different spectral ranges in  different exposures.
There are no measurable temperature changes for the short exposures of
the calibration lamps but during the much longer science exposures the
temperature drifts generally by $\leq 0.2$ K, The estimates for
UVES are of 50 \ms\,  for $\Delta$T = 0.3 K or a $\Delta$P = 1 mbar
\citep{Kaufer2004}, thus assuring a radial velocity
stability within $\sim 50$ \ms.

The motion of Earth during observation smears out the line by $\pm$ 40 \ms,
since the lineshape itself remains symmetric, this does not directly impact
the centroid measurements but it will produce an absorption profile
that is no longer  strictly Gaussian (or Voight) but rather
slightly squared-shaped which further limits the quality of a line fit and must be
considered for multi-component fits of high resolution spectra.

A stronger concern is the possibility of much larger 
distortions within the spectral orders which  have been investigated at the
Keck/HIRES spectrograph by comparing the ThAr wavelength
scale with a second one established from I2-cell observations of a bright
quasar by \citet{Griest10}. They find absolute offsets 
which can be as large as \mbox{500 - 1000 \ms}  and an 
additional distortion of about $300$ \ms\, within the individual orders.

\begin{figure}[b]
\includegraphics[clip=true, width=\linewidth]{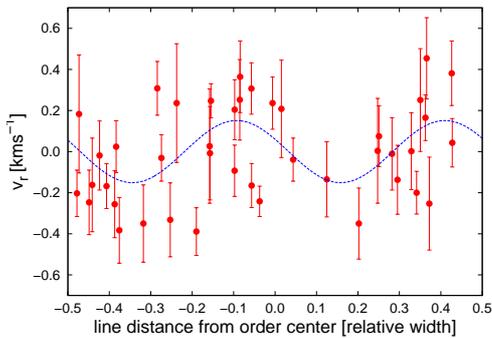}
\caption{All 42 lines with their radial velocity against their
relative position within their order. A cosine fit with an amplitude of 151 \ms
is shown in blue to indicate the possible intra-order distortion.}
\label{fig:v_dwav}
\end{figure}

This would introduce relative velocity shifts between
different absorption features up to a magnitude the analysis
 with regard to $\Delta\mu/\mu$ is sensitive to.
\citet{Whitmore10} repeated the same test for UVES with similar results
though the distortions fortunately show lower peak-to-peak velocity
variations of $\sim 200$ \ms, and \citet{Wendt12} detected 
an early indication of this
effect directly in the measured positions of \hhh features as well as illustrated in
Figure \ref{fig:v_dwav}.
\citet{Molaro11} suggested to use the available solar lines atlas in combination
with high resolution asteroid spectra taken close to the QSO observations to
 check UVES interorder distorsions and published
a revised solar atlas in \citet{Molaro12}.
Such asteroid spectra were used as absolute
calibration to determine velocity drifts in their data via cross-correlation of individual wavelength
intervals in \citet{Rahmani13}.
They found distinct long range drifts of several \mbox{100 \ms} within 1000 \AA.
The origin of these drifts remains currently unknown but is under investigation and was considered
by \citet{Bagdonaite13b} to contribute to their positive signal.
\citet{Rahmani13} found the drift to be constant over a certain epoch and
applied suitable corrections. So far the drift, when present, in UVES spectra always showed the
same trend at different magnitudes which could be an explanation for the reported tendency towards
positive variation in $\mu$ (see next section).

\section{Conclusions and outlook} 
\label{sec:results}
\begin{figure}
\includegraphics[angle=0, width=\linewidth]{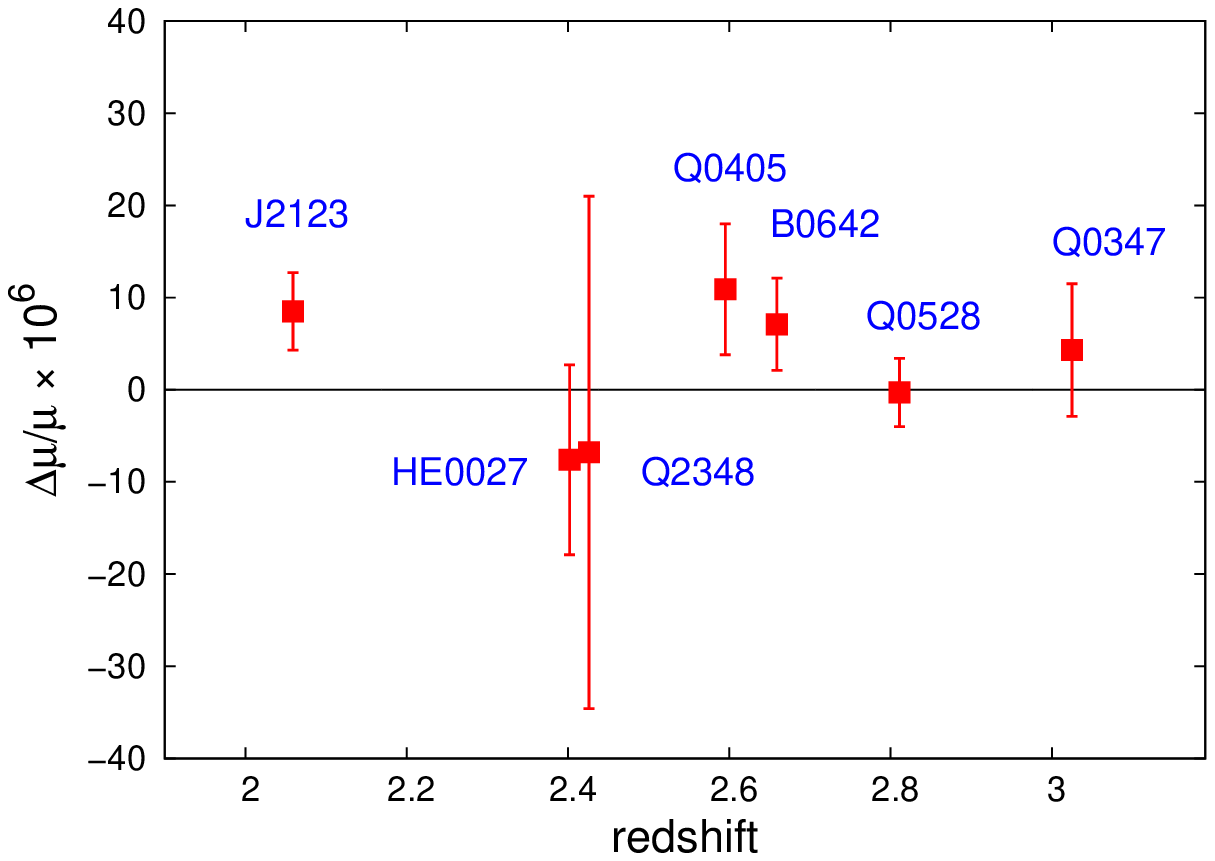}
\caption{Latest  results for \dmu based on \hhh in seven different quasar sightlines observed with UVES:
J2123-0050 \citep{Weerdenburg11}, HE0027-1836 \citep{Rahmani13},
Q2348-011 \citep{Bagdonaite12}, Q0405-443 \citep{King08},
B0642-5038 \citep{Bagdonaite13b}, Q0528-250 \citep{King11},
Q0347-383 \citep{Wendt12}. The given errorbars are the sum of statistical and systematic
 error (if both are given) under the assumption of gaussian distributed errors.}
\label{fig:results}
\end{figure}
Figure \ref{fig:results} shows the latest measurements of \dmu based on \hhh observations with UVES
for 7 observed quasar spectra. The described measurements of QSO 0347-383 in \citet{Wendt12} 
constitute the \dmu measurements via \hhh at the highest redshift to this day.
The presented data of seven measurements yields a mean of \dmu = ( 3.7 $\pm$ 3.5 ) $\times$ 10$^{-6}$ 
and is in good agreement with a non-varying proton-to-electron mass ratio.
Such a generic mean value does not take into account any interpretation with regard to spatial or temporal
 variation and instead merely evaluates the competitive data available for \dmu based on \hhhns-observations, 
which consequently are limited to the redshift range of $2 <$ $z_\mathrm{abs} < 3$ and the evidence
these data provide for any non constant behavior of $\mu$ over redshift.
This tight constraint already falsifies a vast number of proposed theoretical
 models for varying $\mu$ or $\alpha$.
\citet{Thompson2013} come to the conclusion that that ``adherence to the measured invariance in $\mu$ is a
 very significant test of the validity of any proposed cosmology and any new physics it requires''.

The data from the ESO LP observations has the potential to set
 a new cornerstone in the assessment of variability of fundamental physical constants such as $\alpha$
or $\mu$ via measurements of \hhh at high redshifts.
 Observations featuring instruments in the foreseeable
future will provide further insights. Data taken
 with laser-comb calibrated spectrographs (such as CODEX or EXPRESSO)
at large telescopes (E-ELT or VLT, respectively) implicate new methods of data analysis as well.
 
\begin{acknowledgements}
 We are thankful  to the organizers of the
conference on ``Varying fundamental constants and dynamical dark energy'' in Sesto, Italy and 
 express our gratitude to its attendants for numerous fruitful discussions.
We also appreciate the helpful comments by Thorsten Tepper Garc\`ia.
\end{acknowledgements}

\bibliographystyle{aa}

\end{document}